\documentclass[11pt]{article}
\usepackage{moriond,epsfig}

\def\Journal#1#2#3#4{{#1} {\bf #2}, #3 (#4)}
\def\AccJournal#1#2{{#1} (#2)}

\def\NPB{{\em Nucl. Phys.} B}

\def\PRL{\em Phys. Rev. Lett.}
\def\PRD{{\em Phys. Rev.} D}

\def\be{\begin{equation}}
\def\ee{\end{equation}}
\def\bea{\begin{eqnarray}}
\def\eea{\end{eqnarray}}

\begin{document}
\vspace*{4cm}
\title{DIBOSON PRODUCTION CROSS SECTIONS AT $\sqrt{s}=1.96$~TeV}

\author{ J. SEKARIC }

\address{Department of Physics, 513 Keen Building,\\
Florida State University, Tallahassee, Florida, USA}
\address{for the CDF and D{\O} Collaborations}

\maketitle\abstracts{
  The increasing size of the data recorded by the CDF and D{\O} experiments at the Tevatron collider at 
  $\sqrt{s}=$ 1.96~TeV makes the diboson physics program more accessible for probes of the electroweak 
  gauge structure in the Standard Model. Here we summarize the most recent measurements of the diboson 
  cross sections and limits on the trilinear gauge boson couplings.}

\section{Introduction}
  The diboson production plays an important role in electroweak precision measurements and searches for 
  the New Physics (NP) which may exist at some energy scale $\Lambda$. The search for the NP is often 
  related to the precision measurements such those of the cross sections and the trilinear gauge 
  boson couplings (TGCs). More discussions about these TGC parameters can be found in~\cite{tgc}. 
  Depending on the NP scenario, these observables are expected to 
  deviate from their SM predictions. The charged TGCs studied in $WW$ and $W\gamma$ production are 
  $(\Delta{g_{1}^{Z}},\Delta\kappa_{\gamma,Z},\lambda)$ and $(\Delta\kappa_{\gamma},\lambda)$, respectively, 
  and $(\Delta{g_{1}^{Z}},\Delta\kappa_{Z},\lambda)$ in $WZ$ production, where $\Delta$ represents the 
  deviation from the SM prediction. In the SM, $\Delta{g_{1}^{Z}}=\Delta\kappa_{\gamma,Z}=\lambda=$ 0. The 
  neutral TGCs $h_{30,40}^{\gamma,Z}$ studied in $Z\gamma$ production are not allowed in the SM and their 
  values are predicted to be zero. Besides, the diboson production is an important background to studies 
  of the top quark and searches for the Higgs boson and SUSY particles. The most precise knowledge 
  of background processes and their proper modeling is highly valuable for current and future studies.

\section{Diboson Production}

\subsection{$Z\gamma\rightarrow\nu\nu\gamma$}

\begin{figure}[htb] \centering
  \begin{minipage}{16pc}
  \includegraphics[width=17pc]{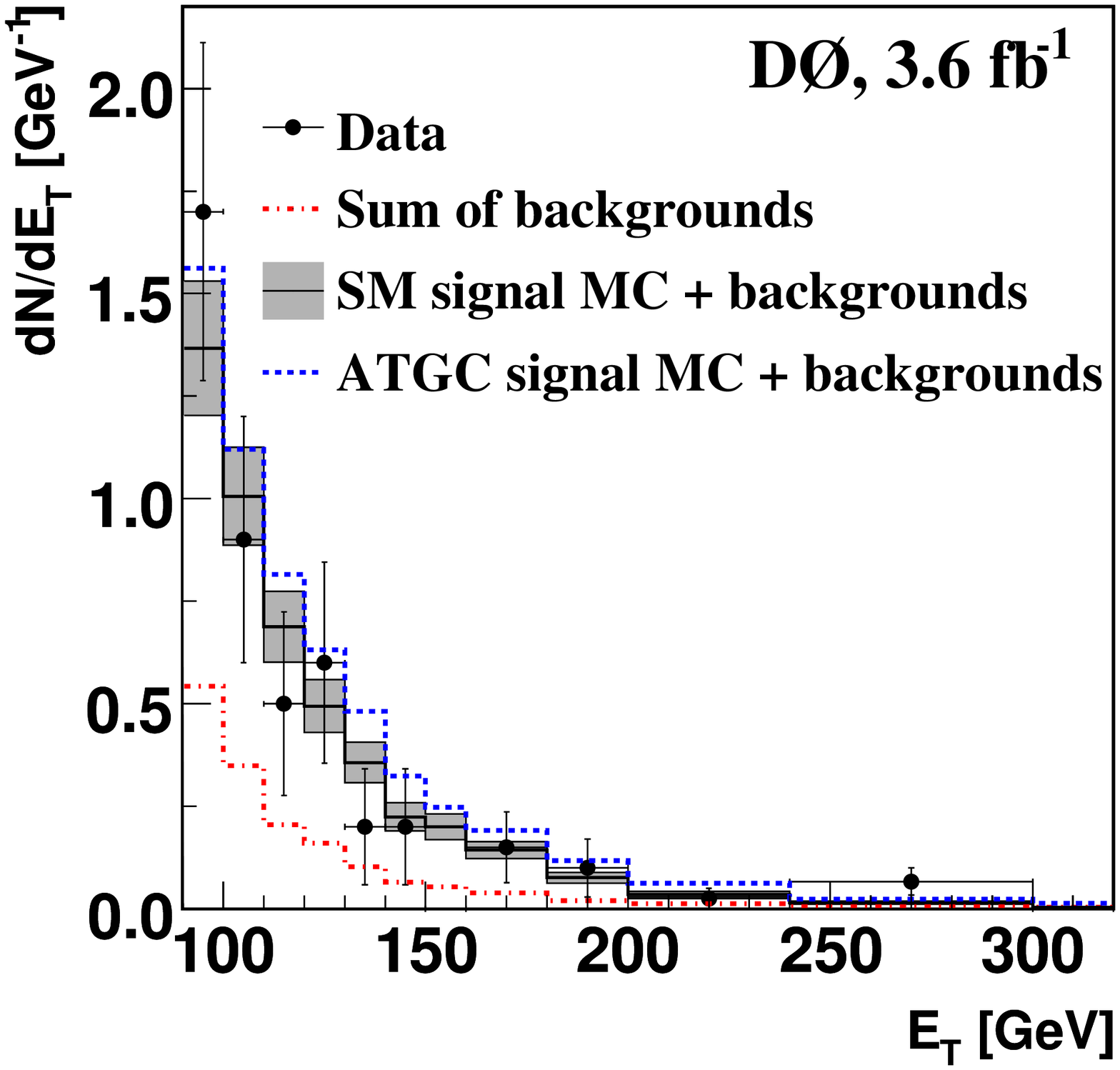}
  \caption{\label{Fig1} Photon $E_{T}$ spectrum of $\nu\nu\gamma$ data candidate
   events compared to the SM signal and background, and the expected distribution
   in the presence of anomalous TGCs. The systematic and statistical uncertainties
   on the SM MC events are included as shaded bands.}
  \end{minipage}\hspace{3pc}
  \begin{minipage}{17pc}
  \includegraphics[width=17pc]{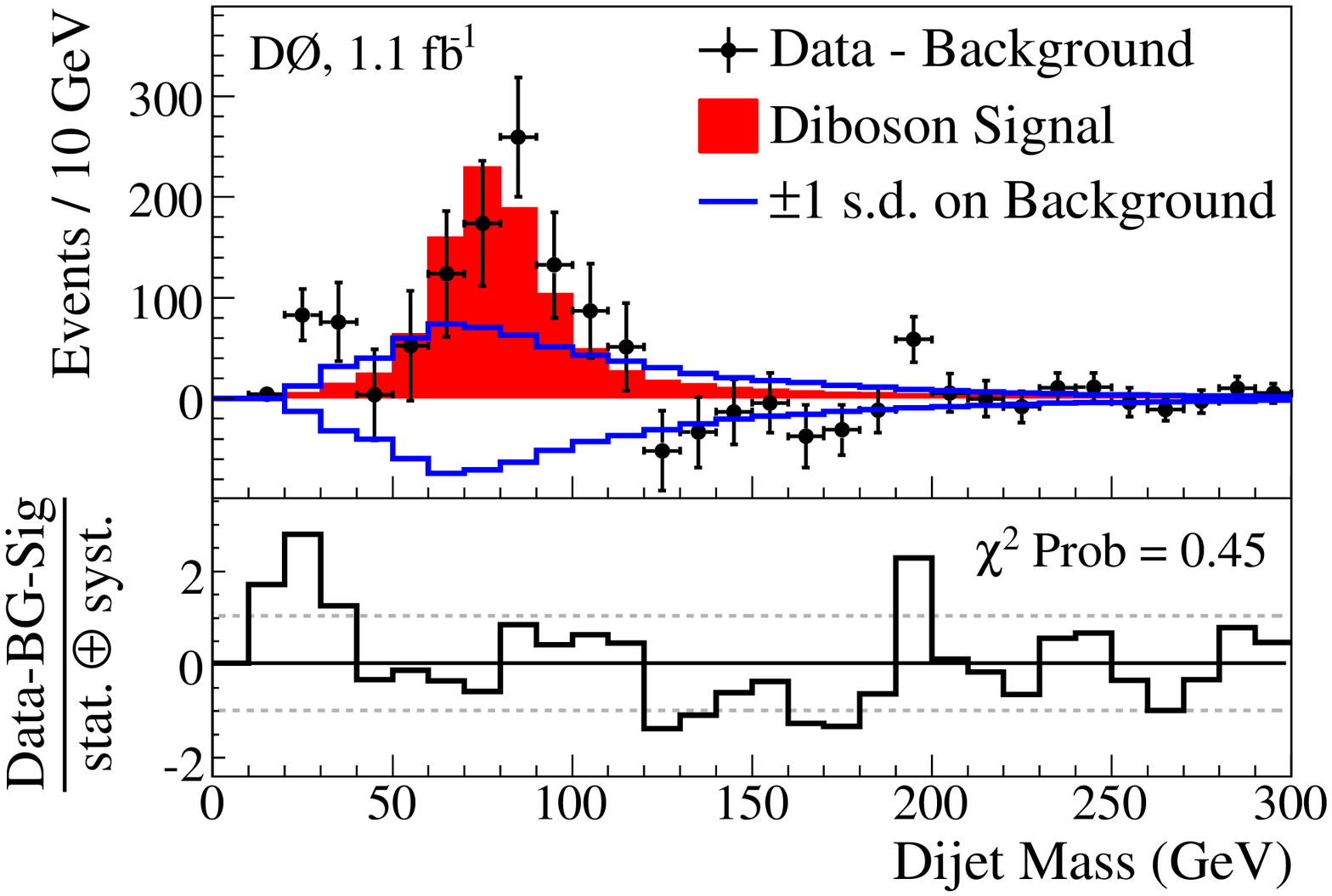}
  \caption{\label{Fig4} A comparison of the extracted signal (filled histogram) to background-subtracted 
    data (points), along with the $\pm1$ standard deviation (s.d.) systematic uncertainty on the background. 
    The residual distance between the data points and the extracted signal, divided by the total uncertainty, 
    is given at the bottom. }
  \end{minipage} \end{figure}

  The $Z\gamma$ events are reconstructed from 3.6 fb$^{-1}$ of D{\O} data. Candidate events are 
  required to have one isolated central photon ($|\eta_{det}|<$ 1.1~\cite{measures}) with $E_{T}>$ 
  90~GeV and $\not{\!\!E_{T}}>70$~GeV. The pointing algorithm~\cite{pointalg} is used in order to 
  reduce the contribution from bremsstrahlung photons. After all selection criteria were applied 51 
  $\nu\nu\gamma$ candidate events are observed. The predicted numbers of signal and background events 
  are $33.7\pm{3.4}$ and $17.3\pm{2.4}$, respectively. The dominant background events are 
  $W\rightarrow{e}\nu$ in which the electron is misidentified as a photon and it contributes with 
  $9.7\pm 0.6$ events. The measured cross section is 
  $\sigma_{ZZ}\times{BR(Z\rightarrow\nu\nu)}=32\pm{9}(stat+syst)\pm{2}(lumi)$ fb~\cite{ZgammaI} which 
  is in agreement with the next-to-leading (NLO) cross section of ($39\pm 4$) fb~\cite{NLOgenZgamma}. 
  The observed signal significance is 5.1 standard deviations (s.d.). Furthermore, the photon $E_{T}$ 
  spectrum shown in~Fig.~\ref{Fig1}, is used to set the limits on $ZZ\gamma$ and $Z\gamma\gamma$ TGCs. 
  The 95$\%$ C.L one-dimensional limits for ${h_{30,40}^{\gamma,Z}}$ at $\Lambda = 1.5$~TeV are 
  $|h_{30}^{\gamma}|<0.036$, $|h_{30}^{Z}|<0.035$ and $|h_{40}^{\gamma,Z}|<0.0019$~\cite{ZgammaI}. The 
  combination with the previous results in the most restrictive limits on these couplings at 95$\%$ C.L. 
  of $|h_{30}^{\gamma,Z}|<0.033$ and $|h_{40}^{\gamma,Z}|<0.0017$ of which three of them 
  ($h_{40}^{\gamma},h_{40}^{Z}$ and $h_{30}^{Z}$) are world's best to date.

\subsection{$ZZ\rightarrow\nu\nu{l^{+}l^{-}}$}

  For the first time, the $ZZ$ production in the $\nu\nu{ll}$ ($l=e,\mu$) final states has been studied 
  at the D{\O} using 2.7 fb$^{-1}$ of data~\cite{llnunu}. The analysis builds up a new variable 
  $\not{\!\!E^{'}_{T}}$, highly discriminating against the $Z\rightarrow{ll}$ background events. Its purpose 
  is to minimize the mismeasurement of the transverse momentum of either the charged leptons or the hadronic 
  recoil system which contributes to the reconstructed $\not{\!\!E_{T}}$. In the electron channel 
  $\not{\!\!E^{'}_{T}}$ is required to be $>27$~GeV, and in the muon channel $\not{\!\!E^{'}_{T}}>30(35)$~GeV for 
  data collected during 2002-2006 (2006-2008). In addition, each channel requires that there 
  are only two oppositely charged isolated leptons with $p_{T}>15$~GeV with the dilepton invariant mass is
  ($70<M_{ll}<110$)~GeV. Events with additional lepton of the same family are vetoed. Further separation 
  between the signal and $WW,~WZ,~W$+jets, $ZZ\rightarrow{l^{+}l^{-}}$ backgrounds uses the likelihood 
  discriminant. The total number of 43 candidate events has been selected of which $26.5\pm{0.5}$ are 
  predicted to be the background. The observed signal significance is 2.6 s.d. and the measured cross 
  section is $\sigma_{ZZ}=2.01\pm{0.93}(stat)\pm{0.29}(syst)$ pb which is consistent with the NLO SM 
  predicted cross section of $1.4\pm{0.1}$ pb~\cite{zztheory}. 
  
\subsection{$ZZ\rightarrow{l^{+}l^{-}l^{'+}l^{'-}}$}

  Sensitivity to single lepton cuts plays an important role when selecting $l^{+}l^{-}l^{'+}l^{'-}$ 
  events ($l,l^{'}=e~or~\mu$). Application of tighter selection criteria such as lepton $p_{T}$, dilepton 
  invariant mass and lepton isolation, relative to the previous D{\O} analysis~\cite{zzevid}, results in the first 
  observation of the $ZZ$ production at a hadron collider. The analysis uses 1.7 fb$^{-1}$ of D{\O} data in which 
  four isolated leptons with $p_{T}>15$~GeV are required to be within $|\eta_{det}|<2$ if muons and 
  $|\eta_{det}|<1.1$ or $1.5<|\eta_{det}|<3.2$ if electrons. Two most energetic leptons in the $\mu\mu\mu\mu$ or 
  $eeee$ channel are required to have $p_{T}>30(25)$~GeV. Events are required to have at least one $Z$ with an 
  invariant mass greater than 70~GeV and the other greater than 50~GeV. In the $ee\mu\mu$ channel the two most 
  energetic electrons or muons must have $p_{T}>25(15)$~GeV. Lepton candidates are required to be spatially 
  separated by $R>0.2$ ($R=\sqrt{(\Delta\eta)^{2}+(\Delta\phi)^{2}}$) to reduce $Z\rightarrow\mu\mu$ background. 
  The total of 3 events, two in the $eeee$ channel and one in the $\mu\mu\mu\mu$ channel, were observed. The SM 
  signal and background are expected to contribute with $1.89\pm{0.08}$ and $0.14^{+0.03}_{-0.02}$ events, 
  respectively. The observed signal significance is $5.3\sigma$ and the measured cross section
  $\sigma_{ZZ}=1.75^{+1.27}_{-0.86}(stat)\pm{0.13}(syst)$ pb~\cite{zzobs} is in agreement with the SM 
  prediction~\cite{zztheory}. The combination with previous analyses~\cite{llnunu} and~\cite{zzevid} results in 
  the observed signal significance of $5.7\sigma$ and the measured cross section is 
  $\sigma_{ZZ}=1.60\pm{0.63}(stat)^{+0.16}_{-0.17}(syst)$ pb.

\subsection{$WZ\rightarrow{jjll}$}

  A search for anomalous $ZWW$ TGCs in the $jjll$ final states results in setting the 95\% C.L. limits on 
  $\Delta\kappa_{Z},\lambda$ and $\Delta{g_{1}^{Z}}$ using the $p_{T}^{ll}$ distribution, and the 95\% C.L. 
  limits on the cross section for the $WZ$ production~\cite{lljj}. This preliminary result is obtained using 
  1.9 fb$^{-1}$ of CDF data selecting events with two high $p_{T}$ leptons and two jets. The $p_{T}^{ll}$
  phase space is divided in three regions: control region used to validate the data modeling ($105-140$~GeV), 
  medium ($140-210$~GeV) and high ($>210$~GeV) regions. The medium and high $p_{T}^{ll}$ regions are used to 
  perform the measurements. Total number of 
  observed (predicted) events is 97 ($71.4\pm{0.5}$) and 12 ($9.74\pm{0.18}$) in the medium and high region, 
  respectively. The observed 95\% C.L. limits on TGCs are $-1.09<\Delta\kappa_{Z}<1.40$, $-0.18<\lambda <0.18$ 
  and $-0.22<\Delta{g_{1}^{Z}}<0.32$ for $\Lambda=1.5$~TeV, and $-1.01<\Delta\kappa_{Z}<1.27$, 
  $-0.16<\lambda{<0.17}$ and $-0.20<\Delta{g_{1}^{Z}}<0.29$ for $\Lambda =2$~TeV. Unbinned MC fit to data in 
  the dijet mass distribution results in the 95\% C.L. cross section limits of 234 fb and 135 fb for medium 
  and high region, respectively.

\subsection{$WW\rightarrow{l}\nu{l'}\nu$}

  The most precise $WW$ cross section measurement at a hadron collider is performed analyzing the 
  $l\nu{l'}\nu$ ($l,l'=e,\mu$) final states with 1.0 fb$^{-1}$ of D{\O} data~\cite{lnulnu}. In each $ll'$ 
  final state ($ee,\mu\mu$ or $e\mu$) the two most energetic leptons are required to have $p_{T}>25~(15)$~GeV, 
  to be of opposite charge and to be spatially separated from each other by $R>0.8$ ($ee$) and $R>0.5$ ($e\mu$). 
  The $Z/\gamma^{*}\rightarrow{ll}$ background is effectively removed requiring $\not{\!\!E_{T}}>45$ ($ee$), 
  20 ($e\mu$) or 35 ($\mu\mu$)~GeV, $\not{\!\!E_{T}}>50$~GeV if $|M_{Z}-m_{ee}|<6$~GeV ($ee$), 
  $\Delta\phi_{\mu\mu}<2.45$ and $\not{\!\!E_{T}}>40$~GeV if $\Delta\phi_{e\mu}>2.8$. Imposing the upper cut 
  on the transverse momentum of the $WW$ system, of 20 ($ee$), 25 ($e\mu$) and 16 ($\mu\mu$)~GeV minimizes 
  the $t\bar{t}$ background. After all selection criteria were applied, all three combined channels yield 
  100 candidate events, $38.19\pm{4.01}$ predicted background events and $64.70\pm{1.12}$ predicted signal 
  events. The cross section measurements in the individual channels are combined, yielding 
  $\sigma_{WW}=11.5\pm{2.1}(stat+syst)\pm{0.7}(lumi)$ pb which is in agreement with the SM NLO prediction of 
  $12.4\pm{0.8}$ pb~\cite{zztheory}. The $p_{T}$ distributions of the leading and trailing leptons were used 
  to set limits on anomalous TGCs considering two different parameterizations between the couplings. The 
  one-dimensional 95\% C.L. limits for $\Lambda=2$~TeV are $-0.54<\Delta\kappa_{\gamma}<0.83$, 
  $-0.14<\lambda_{\gamma}=\lambda_{Z}<0.18$ and $-0.14<\Delta{g_{1}^{Z}}<0.30$ under the 
  $SU(2)_{L}\times{U(1)_{Y}}$-conserving constraints~\cite{su2u1}, and 
  $-0.12<\Delta\kappa_{\gamma}=\Delta\kappa_{Z}<0.35$ and $-0.14<\lambda_{\gamma}=\lambda_{Z}<0.18$ under the 
  assumption that $\gamma{WW}$ and $ZWW$ couplings are equal.
 
\subsection{$WW+WZ\rightarrow{l}\nu{jj}$}

  This analysis results in the first evidence for the $WW/WZ$ production in the $l\nu{jj}$ final states at the 
  hadron collider with 1.1 fb$^{-1}$ of D{\O} data~\cite{lnujj}. Selected $l\nu{jj}$ ($l=e,\mu$) candidate 
  events are required to have a single isolated lepton with $p_T>20$~GeV and $|\eta|<1.1\ (2.0)$ for electrons 
  (muons), $\not{\!\!E_{T}}>20$~GeV and at least two jets with $p_T>20$~GeV. The jet of highest $p_T$ must have 
  $p_T>30$~GeV and the transverse mass of leptonically decaying $W$ boson must be $>35$~GeV to reduce the multijet 
  background. Because of the small signal-to-background ratio ($3\%$), an accurate modeling of the dominant 
  $W$+jets background is essential and therefore, studied in great detail. After all selection criteria were 
  applied, the signal and the backgrounds are further separated using a multivariate classifier, Random Forest 
  (RF)~\cite{SPR}. The signal cross section is determined from a fit of signal and background RF templates to the 
  data with respect to variations in the systematic uncertainties~\cite{poisson} and is measured to be
  $\sigma_{WW+WZ}=20.2\pm{2.5}(stat)\pm{3.6}(syst)\pm{1.2}(lumi)$~pb which is consistent with the SM NLO prediction 
  of $\sigma(WW+WZ)=16.1\pm{0.9}$~pb~\cite{zztheory}. The observed signal significance is 4.4 s.d.. Fig.~\ref{Fig4} 
  shows the $WW/WZ$ dijet mass peak extracted from data compared to the MC prediction. 

\section{Summary}

  The recent results in diboson production at the Tevatron, using $1.0-3.6~$pb$^{-1}$ of data, 
  have been presented. Measured cross sections for the $Z\gamma,ZZ,WZ$ and $WW$ processes and TGCs at 
  the $\gamma{ZZ},\gamma{WW}$ and $ZWW$ vertices are in agreement with the SM predictions. The D{\O}
  experiment sets the world's tightest limits on the $h_{40}^{\gamma},h_{40}^{Z}$ and $h_{30}^{Z}$ TGCs
  and reports the first evidence of $WW/WZ$ production in semi-leptonic final states at a hadron collider.

\begin{thebibliography}{99}
\bibitem{measures}D{\O} uses a cylindrical coordinate system with the $z$ axis running along the  
                  beam axis. Angles $\theta$ and $\phi$ are the polar and azimuthal angles, 
                  respectively. Pseudorapidity is defined as $\eta = -\ln \left[ \tan(\theta/2) \right]$ 
                  in which $\theta$ is measured with respect to the primary vertex. In the massless limit, 
                  $\eta$ is equivalent to the rapidity $y = (1/2)\ln \left[ (E+p_z)/(E-p_z) \right]$. 
                  $\eta_{\rm det}$ is the pseudorapidity measured with respect to the center of the detector.
\end{thebibliography}
\end{document}